# Atomic structure and electronic properties of folded graphene nanoribbons: a first-principles study

Wenjin Yin,[1, 2] Yuee Xie,[1] Li-Min Liu[*2], Yuanping Chen,[*1] Ru-Zhi Wang[3,2], Xiao-Lin Wei[1,2], and Leo Lau[4,2]

[1]Department of Physics and Laboratory for Quantum Engineering and Micro-Nano Energy Technology, Xiangtan University, Xiangtan 411105, Hunan, China
[2]Beijing Computational Science Research Center, Beijing 100084, China
[3]Laboratory of Thin Film Materials, College of Materials Science and Engineering, Beijing University of Technology, Beijing 100124, China
[4] Chengdu Green Energy and Green Manufacturing Technology R&D Center, Chengdu, Sichuan, 610207, China

[*1]Corresponding author: chenyp@xtu.edu.cn
[*2]Corresponding author: limin.liu@csrc.ac.cn

**Abstract**

Folded graphene nanoribbons (FGNRs) have attracted great attention because of extraordinary properties and potential applications. The atomic structure, stacking sequences and electronic structure of FGNRs are investigated by first-principles calculations. It reveals that the common configurations of all FGNRs are racket-like structures including a nanotube-like edge and two flat nanoribbons. Interestingly, the two flat nanoribbons form new stacking styles instead of the most stable AB-stacking sequences for flat zone of bilayer nanoribbons. The final configurations of FGNRs are associated with several factors of initial structures, such as interlayer distance of two nanoribbons, stacking sequences, and edge styles. The stability of folded graphene nanoribbon greatly depends on the length, and it can only be thermodynamically stable when it reaches the critical length (~60 Å). The band gap of the folded zigzag graphene nanoribbons becomes about 0.17 eV, which provides a new way to open the band gap.

## I. Introduction

Graphene a two-dimensional (2-D) crystal has attracted recent tremendous theoretical and experimental attentions due to its intriguing electronic, mechanical, thermal, and optical properties.[1-6] With its extraordinary properties, graphene is highly expected to be the material of novel technological devices.[7,8] Graphene nanoribbons, which are quasi-one-dimensional strips of graphene, can be easily synthesized through lithography and mechanism peeling.[9-11] It turns out that the properties of graphene nanoribbons can be changed from metallic to semiconducting through adjusting the widths and the edges.[12-14]

Recently, folded graphene nanoribbons (FGNRs) have been observed experimentally, which can be achieved via mechanical stimulation, high temperature annealing or irradiated by high energy electrons from graphene nanoribbons.[15-18] The FGNRs can be viewed as two flat graphene nanoribbons continuously connected by a nanotube-like edge. Scanning tunneling microcopy (STM) studies suggested that FGNRs can exist not only in AB-stacking but also in AA-stacking or other patterns.[16,17] The folded graphene nanoribbons possess many exotic electric properties and phenomena, such as zero energy edge states[19], odd and even coexisting conductance steps[20], and metal/half-metal transition[21], etc. Such extraordinary properties of the folded graphene nanoribbons greatly extended the potential application of graphene. Many theoretical and experimental methods to tune physical properties of graphene nanoribbons by folding have also been proposed.[22-24] However, some fundamental properties of the folded graphene nanoribbons, such as the atomic structure, stacking style, the critical width and electronic structure are still unclear until now.

In this paper, folded graphene nanoribbons with both zigzag and armchair edges have been investigated by first principles calculations. Based on the calculations with the different initial configurations, it reveals that after geometrical optimization all FGNRs become racket-like structures including a nanotube-like edge and two flat nanoribbons, and the final stacking styles of the flat nanoribbons not only can be AB-stacking style, but also analogy AB-stacking and even AA-stacking. The geometry of the "racket" and the stacking style is related tightly with the parameters of initial structures. Furthermore, the energy difference between the flat graphene and folded one suggests the folded graphene nanoribbons can only be thermodynamically stable when the width reaches about 61 Å for zigzag and 95 Å for armchair. In addition, it is found that the zigzag FGNRs own a band gap about 0.17 eV, which is different from the Dirac-cone feature of pure graphene.

## II. Method and Structure

The mixed Gaussian and plane-wave basis set code CP2K/QUICKSTEP (QS)[25] has been used for most of the structure calculations. The plane-wave basis set code, VASP, has been used to calculate the band structure.[26] These two codes are both in the framework of the density-functional theory with the generalized gradient approximation (GGA). The energy cutoff is set at 320 Ry for CP2K/QS, and 400 eV for VASP. Perdew-Burke-Ernzerhof (PBE) functional[27] is used for electronic exchange and correlation term. Atoms are fully relaxed until the residual force is less than 0.01 eV/Å. The accuracy of such settings has been recently tested.[28,29] The van der Waals (vdW) interaction is crucial for the formation and stability of the folded graphene nanoribbons, and thus the recently developed DFT-D3 method implemented in CP2K/QS are used in our calculations.[25]

The folded graphene nanoribbon which contains two flat graphene layers and a fractional nanotube in closed edge, which can be formed by folding a single graphene sheet. The final structure of the folded graphene nanoribbons may be affected by the initial structure. The main parameters for the initial structure include that the initial distance between the upper and down flat graphene or the diameter of the nanotube, the stacking sequence between upper and lower layers, and the width of the graphene nanoribbons. To fully explore the final structures of the folded graphene, many different initial configurations of the folded graphene nanoribbons are considered, and all the atomic structures are fully relaxed.

The initial atomic structures of graphene nanoribbons and folded graphene nanoribbons are shown in Fig. 1. Zigzag graphene nanoribbons are represented via the number of C atomic lines $N_Z$ and the width $W_Z$ along its axis (See Fig. 1(a)), and armchair graphene nanoribbons are expressed by $N_A$ and $W_A$, as shown in Fig. 1(b). The symbol $N_\beta / W_\beta$ ($\beta = Z, A$) represents the scale between the number of C atomic lines / the width for both the single graphene nanoribbons and folded graphene nanoribbons as depicted in Fig. 1(a) and (b). The side view of the initial folded graphene nanoribbons was shown in Fig. 1(c), where the symbol D denotes the initial diameter of the nanotube or the distance between the top and bottom layers of the folded graphene nanoribbons. The four different stacking styles of the flat zone: (I) AA-stacking, (II) AB-stacking, (III) AB`-stacking, and (IIII) AB``-stacking are shown in Fig. 1 (e). The top layer, labeled as Layer 1, is colored in blue and the bottom layer, labeled as Layer 2, is colored in light olive brown. A (4×4×1) super-cell is used to simulate the graphene nanoribbons, and the edge atoms of nanoribbons are saturated by hydrogen atoms to avoid the interaction between the upper and lower layer edges. Along the y-axis, the system is in periodic boundary condition (PBC) for the folded graphene nanoribbons, and in the x- and z-axis, about 15 Å vacuum was used to avoid the interaction between the adjacent images.

The relative stability of these final folded graphene nanoribbons in various stacking have been characterized with the formation energy $E_f$, which were calculated through the following equation:

$$E_f = E_t - N_g \varepsilon_g - N_h \varepsilon_h, \quad (1)$$

where $E_t$ is the total energy of the folded graphene nanoribbons, $N_g$ is the total number of C atoms, $\varepsilon_g$ is the energy per C atom, $N_h$ is the total number of hydrogen atoms, and $\varepsilon_h$ is the energy per atom. The smaller formation energy means that the structure is more stable.

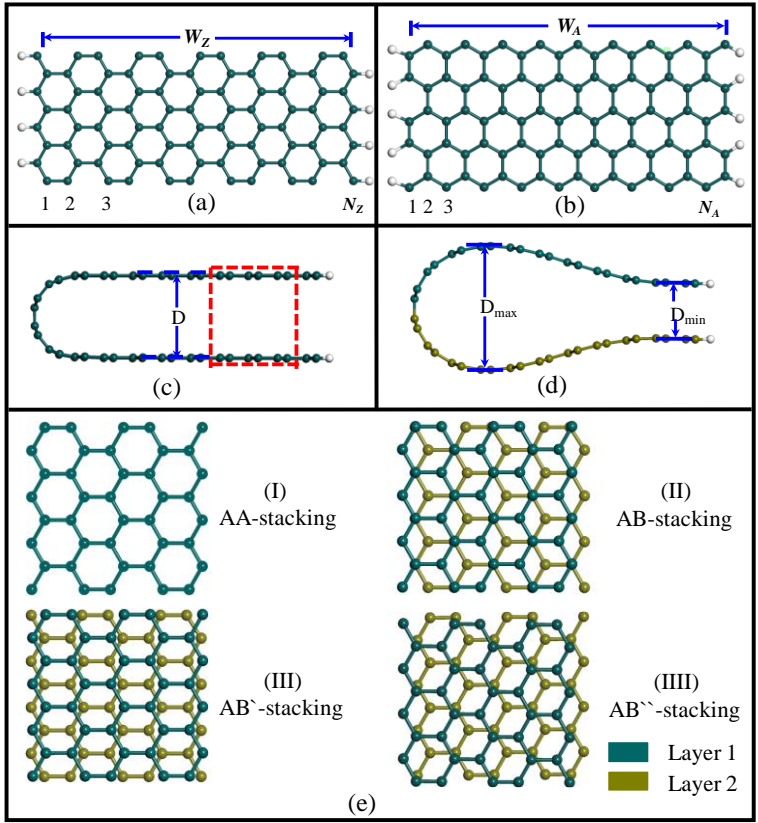

Fig. 1 (Color online) The atomic structures of graphene nanoribbons and an folded graphene nanoribbons. Top view of the perfect zigzag (a) and armchair (b) graphene nanoribbons, and the width of C atomic lines for zigzag and armchair are $N_Z$ and $N_A$, respectively. (c) Side view of the initial folded graphene nanoribbons, which contains a fraction of nanotube in closed edge and two flat sheets in open edges. The symbol D denotes the initial distance between the upper and bottom flat sheets. (d) Side view of the relaxed folded graphene nanoribbons. The symbol $D_{max}$ represents the distance between the maximum distance for the curved part and $D_{min}$ is the distance between the two flat layers. (e) Top view of the four different stacking styles for the folded graphene nanoribbons, and such zone is shown in the red dashed box in (c): (I) AA-stacking, (II) AB-stacking, (III) AB`-stacking, and (IIII) AB``-stacking. The open edges are saturated with hydrogen atoms in white balls, while others are C atoms.

## III. Results and Discussion

As mentioned above, all atomic structures of the folded graphene nanoribbons are fully relaxed from the different initial configurations, such as the different interlayer distance, stacking and so on, in order to explore the final configurations. Interestingly, starting from the different initial configurations, all folded graphene nanoribbons become racket-like structures after geometrical optimization, as shown in Fig. 1(d). The folded graphene nanoribbons can be divided into two parts: One part is a nanotube-like edge, in which the maximum top-bottom separation in the curved region is denoted as $D_{max}$. The other part is two flat nanoribbons, which mimics the bilayer graphene nanoribbons and the layer distance can be denoted as $D_{min}$.

The final geometry and stacking style of the folded structures are greatly affected by the initial structures, such as the interlayer distance of two nanoribbons, edge styles, and width of the nanoribbons. Let us first concentrate on the zigzag folded nanoribbons. As shown in Fig. 2, the final configurations of the zigzag folded nanoribbons can be divided into two types by the number Nz of C chains: one is even, and the other is odd. Considering that Nz=40 and Nz=41 are the typical thermodynamically stable nanoribbons, as discussed below, these two types of nanoribbons are used to discuss the features of odd and even number of C atom chains except we noted. As for the even number of C atomic lines ($N_Z$=40), the final stacking styles of the structures are mainly regulated by the initial distance D. As shown in Fig. 2(a), three different zones based on the initial and final configurations can be seen. When the initial distance D is smaller than 5 Å, all structures arrive at AB-stacking style after the structures relaxation from the different stacking sequences (see dashed line in Fig. 2 (a)). When the initial distance D is in the range of 5 Å ~ 8 Å, the final configurations keep the same stacking sequence for both AA- and AB-stacking styles, while AB`-stacking style transforms into AB-stacking after optimization (see dash dotted line). The final structures remain the initial stacking when the initial distance D is larger than 8 Å (see short dash dotted line). It is well-known that the AB stacking is the most stable one. Recently, scanning tunneling microcopy observed experimentally that folded graphene can not only in AB-stacking but also in AA-stacking or other patterns.[16] Our current calculations explain well such "abnormal" AA-stacking for the folded graphene sheets.

Beside the stacking style, the two separation distances $D_{max}$ and $D_{min}$ are also greatly affected by the initial distance D, as illustrated in Fig. 2(b). When the initial distance D≤7Å, $D_{max}$ is in the range of 8 Å ~ 10 Å, while $D_{min}$ always keeps the same value at about 3.5 Å, which is close to the interlayer distance of bilayer graphene nanoribbons. When the initial distance D is larger than 7Å, the $D_{max}$ abruptly increases to 12.5 Å ~ 14.5 Å, meanwhile the $D_{min}$ also abruptly increases, which is quite close to the initial D.

As shown in Fig. 2(c), when the initial distance D is smaller than 8 Å, the formation energy of AB-stacking is always lower than AA-stacking, thus AB-stacking is more stable than AA-stacking. Interestingly, the formation energy changes a lot at the initial distance of 8 Å. When D is larger than 8 Å, the formation energy is about 3.2 eV larger than that of D≤8, as shown in Fig. 2(c). The final configurations with the D≥8 Å should be meta-stable. This should be that when the distance between the upper and bottom layers is larger than 8 Å, the interlayer vdW interaction is too smaller to keep these layers together.

Next, let us discuss the final configurations for the odd number of nanoribbons of zigzag folded nanoribbons. The final configuration and distance for the odd number of C atomic lines ($N_Z$=41) are shown in Fig. 2(d) and Fig. 2(e), respectively. Different from the case of even number of C atomic lines, when the initial distance D is smaller than 5Å, after geometrical optimization the stacking style becomes a new stacking, i.e., AB``-stacking, instead of AB stacking (see dotted line in Fig. 2(d)). When D are equal to 5.29 Å and 7.18 Å, AA-stacking will keep the initial AA-stacking while others transform into AB``-stacking (see dashed line in Fig. 2(d)). When D is larger than 8 Å, the final stacking remains the initial stacking styles, as illustrated in blue dash dotted line which is similar to the even C atomic lines $N_Z=40$ shown in Fig. 2(a).

The $D_{max}$ and $D_{min}$ of the final structures as a function of initial distance D based on the final stacking are shown in Fig. 2(e). The $D_{max}$ and $D_{min}$ of the odd C atomic lines ($N_Z=41$) exhibit quite similar feature as those of the even C atomic lines ($N_Z=40$) shown in Fig. 2(b). The formation energy for zigzag folded graphene nanoribbons in odd C atomic lines ($N_Z=41$) are shown in Fig. 2(f). Compared with the case of even C atomic lines in Fig. 2(c), it is unexpectedly found that the new stacking (AB``-stacking) has the lowest formation energy than other stacking styles, which means that the AB``-stacking is more energetically favorable than others for odd C atomic lines. In analogy with the case of $N_Z=40$ at initial distance higher than 8 Å, the foramtion energy in magnitude for $N_Z=41$ is almost the same. In summary, the relative stability and final configuration of the zigzag folded graphene nanoribbons are greatly affected by the stacking styles, the initial distance D, and the odd or even number of C atomic lines $N_Z$.

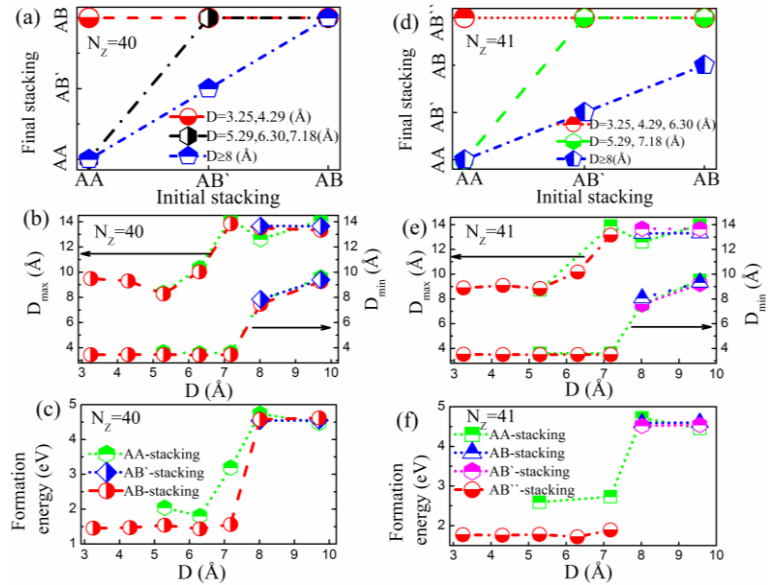

Fig. 2 (color online) The final configurations of zigzag folded graphene nanoribbons as a function of the initial structures. (a), (b), and (c) for the even number of carbon chain with width $N_Z=40$, while (d), (e), and (f) for the odd number of carbon chain with width $N_Z=41$; (a) and (d) show the relationship between the final stacking styles and initial stacking styles in different initial distance D; (b) and (e) exhibit that the maximum distance $D_{max}$ in the bending portion and the minimum distance $D_{min}$ between the upper and bottom layers as a function of the initial distance D; The formation energy of zigzag folded graphene nanoribbons as a function of initial distance D are shown in (c) and (f) for $N_Z=40$ and $N_Z=41$, respectively. The four typical stacking styles AB, AB, AB`, and AB``, are considered in our calculations.

The evolution of the final armchair folded structures as a function of the initial distance D is also investigated, as shown Fig. 3. Different from zigzag folded configurations, armchair structures can only form AA- and AB`-stacking and they keep the initial stacking sequences after fully relaxation. Compared with the zigzag folded graphene nanoribbons, $D_{max}$ and $D_{min}$ of the armchair folded graphene nanoribbons are only regulated by the initial distance rather than stacking styles, as shown in Fig. 3(a). Meanwhile, the values of the $D_{max}$ and $D_{min}$ as a function of initial

D are quite close to those of zigzag folded graphene nanoribbons. In addition, the formation energy for both AA- and AB`-stacking are calculated in relation to the initial distance D, as shown in Fig. 3(b). When the initial distance is smaller than 8 Å, AB`-stacking has lower energy than AA-stacking, indicating that AB`-stacking is more favorable than AA-stacking. It should be noted that the formation jumps when the initial distance is larger than 8 Å, thus the structure with the initial distance > 8 Å should be meta-stable one.

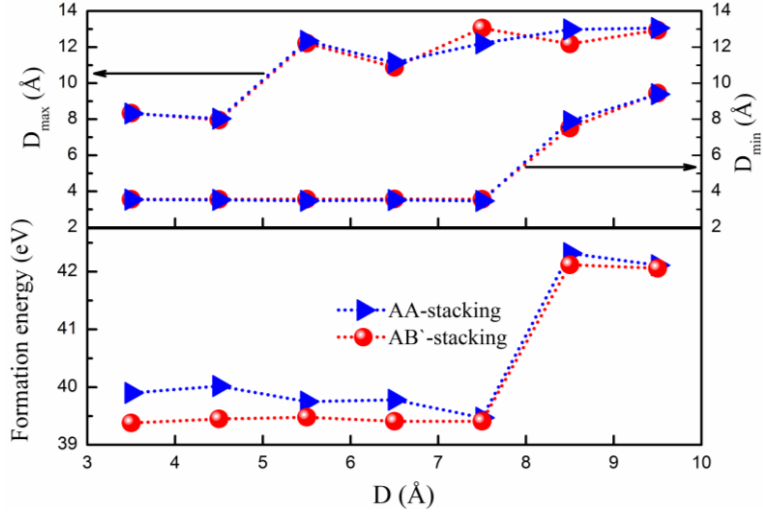

Fig. 3 (Color online) The final structures and the formation energy of armchair folded graphene nanoribbons with various different initial distances, D. (a) The final structures of the $D_{max}$ (left-axis) and $D_{min}$ (right-axis) for the AA- and AB`-stacking in different initial distances, D; (b) The formation energy for the AA- and AB`-stacking as a function of initial distance D. AA-stacking in red dashed line and AB`-stacking in blue short dashed line.

All the results mentioned above is about the relatively stability of the folded graphene nanoribbons compared with the pure graphene. Here, in order to clearly understand the thermodynamic stability of the folded graphene nanoribbons, the energy difference between the folded graphene nanoribbons and the flat ones with the same width are calculated via the equitation written as:

$$\Delta E = E_{fold} - E_{flat}, \qquad (2)$$

where $E_{fold}$ is the total energy of the folded graphene nanoribbons, and $E_{flat}$ is the total energy for the flat graphene with the same C atomic lines as the folded one. When $\Delta E$ is positive, the flat graphene nanoribbons is more stable; When $\Delta E$ is negative, the folded one is more energetically stable.

The calculated energy difference $\Delta E$ as a function of C atomic lines $N$ is shown in Fig. 4. Energy differences of the zigzag and armchair graphene nanoribbons are shown in Fig. 4 (a) and (b), respectively. As shown in Fig. 4(a), the energy difference $\Delta E$ for the zigzag graphene nanoribbons exhibits oscillation as the width increases. It is interesting to note that the magnitude of energy difference is inversely proportional to the odd $N_Z$ or even $N_Z$. By calculating the slope of the dotted line with odd or even $N_Z$, we can obtain the vdW energy is about 24 meV per C atom in parallel flat-like portion, which is a litter smaller than the value 42 meV/atom with the empirical potential simulations[30]. Furthermore, the balance width between the folded and flat graphene is $N_Z=40$ ($W_Z=56.8$ Å) / $N_Z=43$ ($W_Z=61.06$ Å) for the even / odd $N_Z$ zigzag graphene nanoribbons as the energy difference is equal to zero, as shown in Fig. 4(a). Compared with the zigzag folded graphene nanoribbons, the balance width for armchair graphene nanoribbons is at the length with $N_A=78$ ($W_A=94.71$ Å), as illustrated in Fig. 4(b). The difference between zigzag and armchair graphene nanoribbons are mainly resulting from the lattice registry in the bending part. The minimum balance width for zigzag folded graphene nanoribbons in different $N_Z$ should mainly result from the arrangement of end edge atoms. Except for the balance width, the criterion length is 20 Å for the parallel flat-like portion, which agrees with the previous calculation value of 16.2 Å based on atomic-scale finite element method.[17]

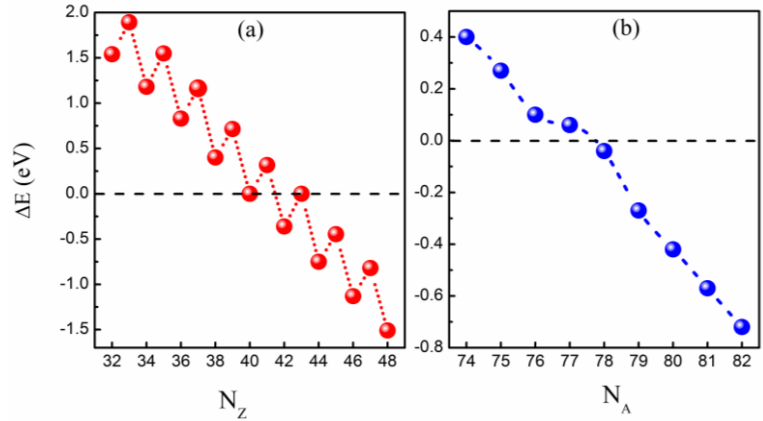

Fig. 4 (Color online) (a) (b)The energy difference between the folded graphene nanoribbons and the flat ones in the same width versus the C atomic lines for the zigzag and armchair edge graphene nanoribbons, respectively.

As mentioned above, the folded graphene can only be thermodynamically stable when it reached the critical size. The recent DFT studies show that the folded armchair graphene can open a small band gap of 0.301 eV for N=40.[30] In order to check both the folding and critical size effects for the folded zigzag graphene, the band structures for several typical configurations are calculated, which are shown in Fig. 5 with (a) $N_Z=20$, (b) $N_Z=21$, (c) $N_Z=40$, and $N_Z=41$. It is well-known that there is no band gap for the perfect graphene. However, it is interesting to find that the folded one has a band gap at $\Gamma$ point. The magnitude of the band gap is 0.05 eV and 0.25 eV for $N_Z=20$ and $N_Z=21$ shown in Fig. 5(a) and Fig. 5(b), and the value of the band gap is about 0.17 eV for both odd ($N_Z=40$) and even ($N_Z=41$) number of C chain in Fig. 5(c) and Fig. 5(d), indicating that the configuration becomes semiconductor. Therefore, energy band can be modified through folding, which provides a method to open the band gap for the developing of electronic device.

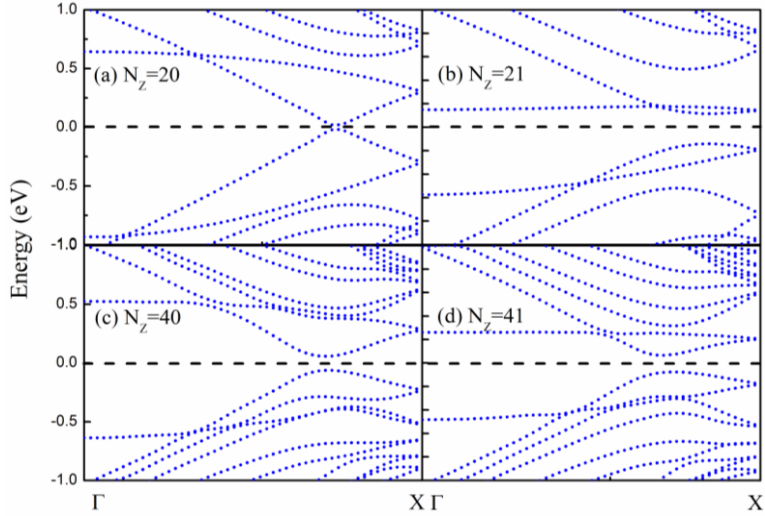

Fig. 5 (Color online) The electronic structure of the zigzag folded graphene nanoribbons. (a), (b), (c), and (d) show the band structures for $N_Z=20$, $N_Z=21$, $N_Z=40$, and $N_Z=41$, respectively.

## IV. Conclusions

The stable atomic structure, stacking sequence, and electronic structure of the folded graphene nanoribbons are investigated by first principles calculations. The calculations show the stability of the folded graphene nanoribbons are regulated by the initial distance, edge styles, stacking styles and the number of C atomic lines. For both the zigzag and armchair folded graphene nanoribbons, the upper and lower layers of the flat zone can form analogue AB stacking (AB`-stacking and AB``-stacking) instead of the most stable AB-stacking sequences as graphite. The calculated the energy differences between the flat and folded graphene suggested that folded graphene is only thermodynamically stable when the equilibrium width reaches $N_Z=40$ ($W_Z=56.8$Å) / $N_Z=43$ ($W_Z=61.06$Å) for even and odd $N_Z$ zigzag folded graphene nanoribbons and the critical width is about $N_A=78$ ($W_A=94.71$Å) for armchair graphene nanoribbons. Upon folding, the band gap can be opened, which provides a feasible way for designing of electronic devices.

**Acknowledgments**

This work was supported by the National Natural Science Foundation of China (Nos. 51006086, 11074213, 51176161, 11244001, and 51032002), the CAEP foundation (Grant No. 2012B0302052), the MOST of China (973 Project, Grant NO. 2011CB922200). The computations supports from Informalization Construction Project of Chinese Academy of Sciences during the 11th Five-Year Plan Period (No.INFO-115-B01) are also highly acknowledged.